\documentclass[]{mn2e} %
\usepackage{graphicx}	
\usepackage{amsmath}	
\usepackage{amssymb}	
\usepackage{bm}    

\def\apj{ApJ} \def\apjs{ApJ. Suppl.} \def\apjl{ApJ. L.} \def\aap{A\&Ap}  \def\mnras{MNRAS} \def\pasj{PASJ} \def\araa{ARA\&Ap}\def\apss{Ap\&SpS}

\title[Molecular Fraction in the Galactic Center]{Molecular Fraction in the Galactic Center: The Central Molecular and HI Zones}

\author[Y. Sofue]{Yoshiaki Sofue\thanks{E-mail: sofue@ioa.s.u-tokyo.ac.jp} \\ 
Institute of Astronomy, The University of Tokyo, Mitaka, Tokyo 181-0015, Japan }

\date{Accepted; Received YYY; in original form} 
\pubyear{2022}
  
\def\vlsr{V_{\rm LSR}}  
\def\vrot{V_{\rm rot}}     
 \def\deg{^\circ} \def\Tb{T_{\rm b}}
\def\Tco{{T_{\rm b}}_{\rm : CO}}
\def\Thi{{T_{\rm b}}_{\rm : HI}}
\def\Ico{I_{\rm CO}}  \def\Ihi{I_{\rm HI}} \def\be{\begin{equation}} \def\ee{\end{equation}} 
 
\def\Xhi{X_{\rm HI}} \def\Xco{X_{\rm CO}}
\def\kms{km s$^{-1}$} \def\Kkms{K km s$^{-1}$}
\def\ekms{{\rm ~km~s^{-1}~}} \def\epc{{\rm ~pc~}}  
\def\be{\begin{equation}} \def\ee{\end{equation}} 
 \def\twco{$^{12}$CO }
\def\sin{\rm {~sin~}}
\def\fmol{f_{\rm mol}}
\def\fsig{f_{\rm mol}^\Sigma}
\def\frho{f_{\rm mol}^\rho} 
\def\Htwo{H$_2$}
\def\xcounit{\Htwo\ cm$^{-2}$ [\Kkms]$^{-1}$}

\def\vcut{V_{\rm cut}} \def\vrot{V_{\rm rot}}
\def\Mc{McClure-Griffiths}
\def\vabs{|\vlsr|}
\def\dev{DEV50}

\def\red{\textcolor{red}}
\def\blue{\textcolor{blue}}
\def\red{}
\def\blue{}

\begin{document} 
\maketitle  
\begin{abstract} 
By mapping the molecular fraction of the Galactic Center (GC), we quantitatively address the question of how much molecular and central the CMZ (Central Molecular Zone) is.
For this purpose we analyze the CO and HI-line archival data, and determine the column- (surface-) and volume-molecular fractions, $\fsig$ and $\frho$, 
\blue{which are the ratio of column-mass density of \Htwo\ projected on the sky to that of total gas (\Htwo\ + HI) from the line intensities, and the ratio of volume-mass densities of \Htwo\ to total gas from the brightness temperature, respectively.}
It is shown that $\fsig$ is as high as $ \sim 0.9-0.95$ in the CMZ, and $\frho$ is $0.93-0.98$ in the GC Arms I and II attaining the highest value of $\sim 0.98$ toward Sgr B2.
The expanding molecular ring (EMR, or the parallelogram) has a slightly smaller $\frho$ as $\sim 0.9-0.93$.
We define the CMZ as the region with $\fsig \ge 0.8-0.9$ between the shoulders of plateau-like  distribution of \Htwo column density from $l=-1\deg.1$ to $+1\deg.8$ having Gaussian vertical distribution with a half thickness of $\pm 0\deg.2$.
\red{The CMZ is embedded in the Central HI Zone (CHZ), which is defined as an HI disc between $l\sim -2\deg$ and $+2\deg.5$, $b=-0\deg.5$ and $+0\deg.5$.}
Based on the analysis, we discuss the origin of CMZ and interstellar physics such as the volume filling factors of molecular and HI gases inferred from the difference between $ \fsig $ and $ \frho $.
\end{abstract}
 

\section{INTRODUCTION} 

The Central Molecular Zone (CMZ) was introduced as the "strong concentration of gas of about 200-pc radius" around the Galactic Center (GC) \cite{morris+1996}.
Decades later today, the exact range of the CMZ is still not well defined, referred to as "a ring-like accumulation of molecular gas in the innermost few hundred parsecs of the Milky Way" \cite{henshaw+2022}.
Some authors think it even smaller area \cite{tsuboi+1999,tokuyama+2019}. 

The purpose of this paper is to quantitatively address the question, how much molecular and central the CMZ is, and to quantitatively define the CMZ.
For such a purpose, the molecular fraction, $\fmol$, defined by the ratio of the molecular hydrogen (\Htwo\ gas) density to the total gas density (HI and \Htwo\ gases) provides the most direct information along with various constraints on the physical and chemical condition of the ISM (interstellar medium)  \cite{elme1993}. 
Since it is a hybrid measure of the gas pressure (density and temperature), metallicity (dust abundance) and UV (ultra violet) radiation pressure, the fraction may be used to determine one of these quantities, given the other two are fixed.
\red{As a first approximation to the discussion of the relative variation of molecular fraction in the GC region, we take the conservative view that the conversion factors from the CO and HI lines to the column densities of \Htwo and H atoms are constant.}

Analyses of large-scale distribution of the molecular fraction in nearby spiral galaxies have been extensively obtained by comparing CO and HI line mapping data, which generally indicate monotonic increase of $\fmol$ toward the galactic centers
\cite{sofue+1995fmol,honma+1995,kuno+1995,hidaka+2002,tanaka+2014}.
More resolved behavior has been obtained in the Milky Way, where the molecular fraction becomes as high as $\sim 0.9$ to unity inside several hundred pc of the Galactic Center \cite{sofue+2016,koda+2016,sofue2022}.
Three dimensional analysis of the CO and HI discs of the Milky Way has also shown a steep vertical increase and saturation of the molecular fraction toward the galactic plane \cite{imamura+1997,sofue+2016}.
The high molecular fraction in the central region seems ubiquitous in spiral galaxies, except for M33's center showing values as low as $\fmol\sim 0.2-0.3$ \cite{tosaki+2011}.
 
In this paper, we investigate the three dimensional distribution of the molecular fraction in the central several degrees of the Milky Way.
We measure the extent and size of the high-$\fmol$ saturated region, and re-define the CMZ using the quantified parameters. 
We assume that the GC distance of the Sun is 8.18 kpc \cite{gravity+2019}. 
 
\section{CO and HI maps}

\subsection{Data}
We make use of the CO- and HI-line data from the archives as follows.
The wide field \twco data were taken from the Columbia Galactic plane CO survey \cite{dame+2001},
\red{which had angular and spectral resolutions of $8'.5$ and 1.3 \kms. respectively}. 
The all-sky HI data were taken from the survey using the Bonn 100-m and Parkes 65-m telescopes \cite{HI4PI+2016}, 
\red{which had effective resolutions of $16'$ and $1.5$ \kms.}
The \twco data for the Galactic Center were taken from the survey of the central $\sim 4\deg \times 1\deg$ region using the 45 m telescope \cite{oka+1998}, which had an effective angular resolution of $37''$ ($={\rm (beam~width^2+grid~spacing^2})^{1/2}$, velocity resolution $2$ \kms, rms noise of $\sim 1$ K, presented by a data cube with grid spacing of $(34'',34'',2 \ekms)$. 
The HI data for the GC region were taken from the survey with the Australia Telescope Compact Array (ATCA) of the central $10\deg\times 10\deg$ region \cite{mcclure+2012}, which had an angular resolution of $145''$, velocity resolution of $1 \ekms$, rms noise of 0.1 K, and are presented in a cube of spacing $(35'',35'', 1 \ekms)$.  

\subsection{Fore- and background disc elimination by velocity cut (DEV)}

The integrated intensity of radio line emission from the GC region around $l\sim 0\deg$ is significantly contaminated by the emission by the fore- and background Galactic disc rotating in the transverse direction to the line sight, where the radial velocity is degenerated  to 'local velocities' at  $|\vlsr|\le \sim 50 \ekms$.
The fore- and background intensity of CO line amounts to the same order as that from the GC disc, and HI emission is stronger than the GC emission by an order of magnitude.

This makes it difficult to measure such quantities properly arising from the GC, particularly in the HI line.
The difficulty can be eased by applying the 'disc-elimination by velocity cut' (DEV) method as used in our previous paper for a study of the three-dimensional structure of the central molecular zone (CMZ) \cite{sofue2022}.
The method simply eliminates low-velocity emission from the integration, or uses only the emission at radial velocities higher than a threshold value, $|\vabs|>\vcut \ekms$.
An obtained map will be called 'DEV50' map, if $\vcut=50 \ekms$.

In this paper, we adopt $\vcut=50$ \kms as read from observed longitude-velocity diagrams (LVD) of the CO and HI line emissions, where the fore- and background disc components are seen as numerous bright horizontal belts in the $(l,\vlsr)$ plane (see the lVD in the later section).

By the DEV procedure, the emission from a fan-shaped area near the $Y$ axis (Sun-GC line) between $\vlsr = \pm 50 \ekms$ on the radial velocity field are eliminated from the integration.
The eliminated-fan angle is $\theta\sim \pm 15\deg$ for $\vabs=50$ and $\vrot\sim 200 \ekms$, which means that about $\sim 15-20\%$ of the whole galactic disc is not counted.
As this also underestimates the emission from the GC disc itself with the same proportion, the total mass of the GC disc may be obtained by multiplying a factor of $\sim 1.2$ to the calculated mass from the DEV map. 

Fortunately, this correction is not needed to derive the molecular fraction, because it contains only the ratio of HI and CO intensities, so that the over-elimination (underestimation) of both components cancels out.

\subsection{DEV50 in the inner Milky Way} 

\begin{figure*} 
\begin{center}     
\includegraphics[width=16cm]{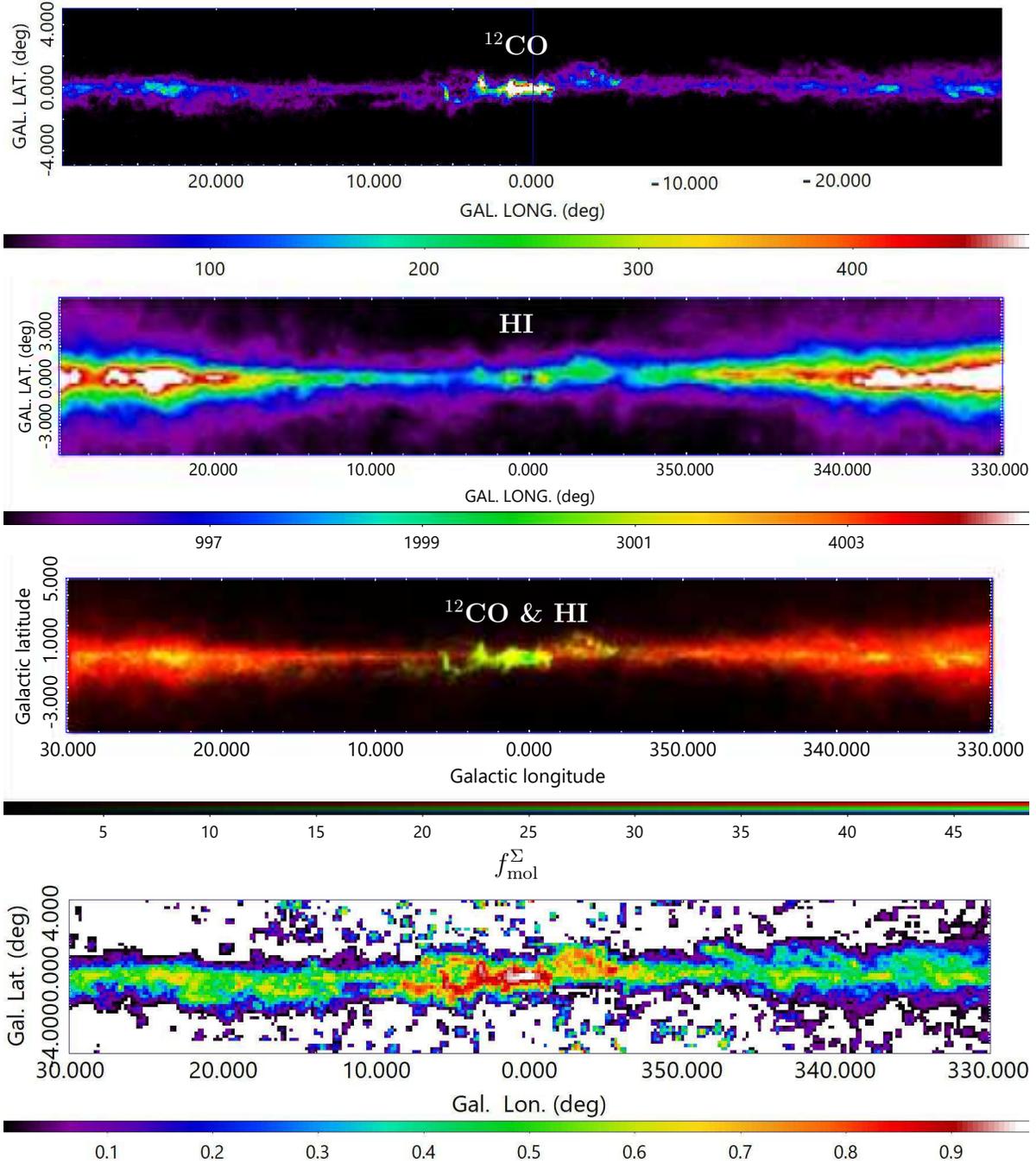}  
\end{center}
\caption{[Top] DEV50 map (fore- and background disc-eliminated intensity map integrate
d over $|\vlsr|\ge 50\ekms$) of \twco line intensity (moment 0) made from the Columbia CO line survey (Dame et al. 2001) in \Kkms.
[2nd] Same, but using the HI4PI HI survey (HI4PI et al. 2016) in \Kkms.
[3rd] Overlay of the HI (red) and CO (green) intensities. 
[Bottom] Molecular fraction made from the DEV50 CO and HI intensity maps.}
\label{mapwide}
\end{figure*}

In Fig. \ref{mapwide} we show the DEV50 ($\vcut=50 \ekms$) maps of the \twco (top panel) and HI (2nd panel) line intensities using cube data from the Columbia CO survey \cite{dame+2001} and HI4PI HI survey \cite{HI4PI+2016}.
\red{In Fig. \ref{mapA}, we show the original, subtracted and DEV50 maps, as reproduced from our previous paper \cite{sofue2022}.}
The 3rd panel shows an overlay of the HI (red) and CO (green) intensities.
The bottom panel shows the molecular fraction, as will be defined and discussed in detail in the later section. 
It is stressed that the intensity maps are quite different from those integrated over the whole velocity range as often used in the literature. 

From the maps we find that:\\
(i) Broadly extended fore- and background disc emissions are not seen.\\
(ii) The central concentration is evident not only in CO, but also in HI.\\
(iii) The GC is surrounded by a less bright or almost empty region at $|l|=\sim 5\deg$ to $20\deg$.\\
(iv) The molecular fraction increases steeply in the central $\sim 5\deg$ making a high-$\fmol$ disc in the GC, confirming the results in the literature \cite{koda+2016,sofue+2016}.

\subsection{DEV50 in the GC}

\begin{figure*} 
\begin{center}   
\includegraphics[width=12cm]{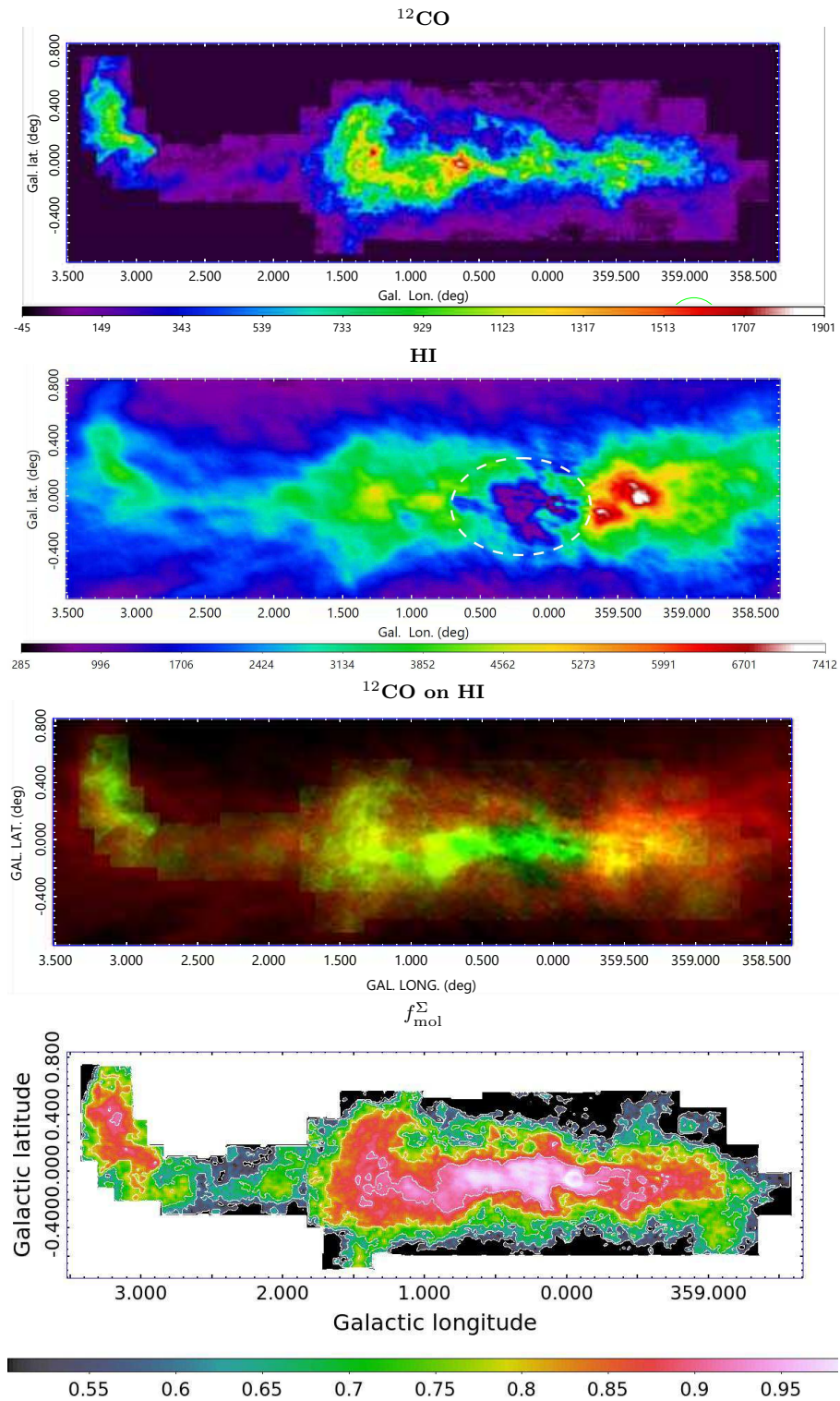} 
\end{center} 
\caption{[Top] DEV50 map of \twco line intensity (moment 0, in \Kkms) integrated over $|\vlsr|\ge 50 \ekms$ made from Nobeyama CO-line survey by Oka et al. (1998).
[2nd panel] DEV50 HI map made from ATCA HI survey by McClure-Griffiths et al. (2012). The dashed ellipse encloses the region with significant absorption against the continuum, and is not used in the analysis.
[3rd panel] Superposition of HI and CO intensity maps as above in red and green colors, respectively.
[Bottom] Molecular fraction made from the above maps. Contours are drawn from $\fsig=0.5$ at interval of 0.05.  }
\label{maps}
\end{figure*}

Figure \ref{maps} presents DEV50 intensity maps of \twco (top panel), HI (2nd panel) and their overlay (3rd panel) in the GC at higher angular resolutions obtained using the cube data from the Nobeyama CO \cite{oka+1998} and ATCA HI \cite{mcclure+2012} surveys. 
The bottom panel shows the molecular fraction obtained from these maps.
Strongly absorbed regions appearing in the HI map near Sgr A and Sgr B against the continuum emission will be avoided in the analysis of the molecular fraction.

The DEV50 CO and HI maps are different from those so far published by integration over the entire velocity range in the sense that:\\
(i) Extended structures have been removed, which is about the same order as the CO intensity and about ten times in HI, so that the structures properly belonging to the GC are clearly distinguishable.\\
(ii) The CMZ ($l=-1\deg.1$ to $1\deg.8$) and CHZ (central HI zone, $l=-2\deg$ to $+2\deg.5$) \cite{sofue2022} are clearly visible.\\
(iii) The overlay demonstrates that the CMZ is embedded in CHZ as usual in disc galaxies.\\
(iv) The highest $\fsig$ region is coincident with the CMZ. The fraction is as high as $\sim 0.9-0.95$ and attains the maximum of $\sim 0.98$ toward Sgr B2. Note that the locally high $\fsig$ clumps near Sgr A are due to HI-line absorption against the radio continuum.

 \subsection{Longitude and latitude profiles}

\begin{figure*} 
\begin{center} 
\includegraphics[width=15cm]{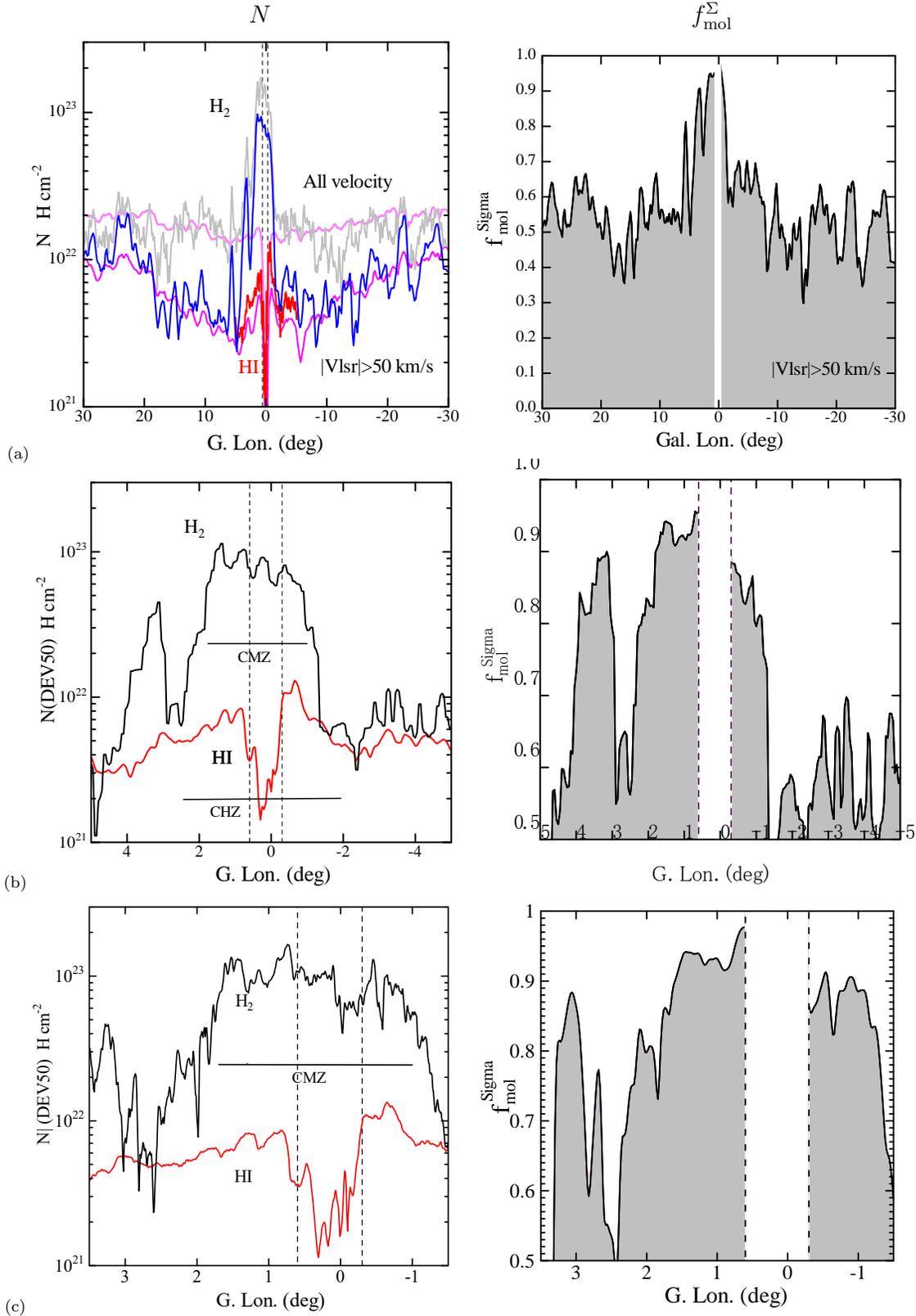} 
\end{center} 
\caption{Longitudinal variation of $N$ (left panels) along the Galactic plane and $\fsig$ (left).
(a) Wide area Columbia CO and HI4PI HI surveys (Fig. \ref{mapwide}). Grey and pink thin lines indicate those without disc elimination for comparison.
(b) Same, but for the central region using Columbia and ATCA HI (\Mc et al. 2012) surveys. 
(c) Same, but using the Nobeyama CO survey (Oka et al. 1998) and ATCA after Fig. \ref{maps}.}  
\label{loncut} 
\end{figure*}

In order to present the characteristic variation of the intensities and molecular fraction more quantitatively, we show in Fig. \ref{loncut} longitudinal profiles of the column density $N$ of H atoms (2H per \Htwo) as calculated from the maps in Fig. \ref{mapwide} and \ref{maps} along the Galactic plane.
The conversion factors are taken constant, as described later.

The top-left panel (a) shows the wide area variations, where thick blue and magenta lines show $N$ from CO and HI intensities after \dev, respectively.
The thin grey and pink lines are the same, but before DEV is applied, so that they include emissions in the whole velocity range.
Difference between DEV and non-DEV profiles is evident.
The non-DEV profiels both in CO and HI (thin lines) are almost flat from $l\sim -30\deg$ to $+30\deg$, except for the central few degrees near CMZ.
On the other hand, the DEV profiles show global minima near the GC an order of magnitude less than those by non-DEV.
The difference is therefore attributed to the 'contamination' of the fore- and background disc components of CO and HI gases in the non-DEV profiles.
 
The CMZ is visible both in DEV and non-DEV profiles as the sharp concentration in the central $-1\deg$ to $+2\deg$ of \Htwo gas.
However, we must be careful that the CMZ's column density (intensity) is overestimated by about $\sim 50\%$, if the DEV is not applied.  

\red{Although far outside of the GC, hence not a main subject here, the correction is still two times at $l\sim \pm 30\deg$.
It is attributed to the still large ratio of the line-of-sight depths inside to outside of the 50-\kms butterfly in Fig. \ref{model}. 
It is interesting to point out that two CO peaks appear at $l\sim 23\deg$ and $-23\deg$ deg in the DEV profile (blue line in Fig. \ref{loncut}(a)), but they not recognized in the non-DEV (grey). This may indicate the usefulness of the method in deriving the Galactic structures from the integrated intensities.}

Panels (b) and (c) show close up of the central region after \dev using the Nobeyama CO survey and ATCA HI survey as well.
The CMZ exhibits its plateaued distribution both in the Columbia and Nobeyama CO data.
The right-hand side panels show the longitudinal distributions of column-molecular fraction $\fsig$, which will be described in detail in the later section.

On the other hand, the HI distribution is milder both in the longitude and latitude directions.
The column-density increases gradually toward the GC, although its peak is significantly disturbed by the absorption against Sgr A and B.
Thus, the CHZ is a concentration without clear cut shoulders, and the $e$ folding radius is measured to be $\sim \pm 2\deg - 2\deg.5$.

\section{Molecular Fraction}

\subsection{Column molecular fraction $\fsig$ from the integrated intensity}

Because the brightness temperature of HI emission at $|\vlsr|\ge 50\ \ekms$ is observed to be less than $\sim 30$ K in the analyzed region (see the longitude-velocity diagrams in the later subsection), while the excitation temperature is considered to be $T_{\rm ex}\sim 140$ K in the disc and still higher in the GC \cite{sofue2018,pohl+2022}, we here assume that the HI line is optically thin.
This makes it simple to convert the intensity to the column density of HI gas.
As to the CO line, which is usually optical thick, we make use of the empirical conversion relation as described below. 

The molecular fraction for the column mass density projected on the sky, hereafter column molecular fraction, is defined by  
\begin{equation}
\fsig
\red{=\frac{\Sigma_{\rm H_2}}{\Sigma_{\rm HI} + \Sigma_{\rm H_2}} }
=\frac{1}{1+\eta_\Sigma}.
\end{equation}
Here, $\eta_i$ is the ratio of column density of HI gas, $\Sigma_{\rm HI}$ (or column number density $N_{\rm HI}$) to that of \Htwo\ gas, $\Sigma_{\rm H_2}$ (or $2N_{\rm H_2}$), which is related to the ratio of HI to \twco line intensities using the conversion factors as 
\be
\eta_\Sigma
=\frac{\Sigma_{\rm HI}}{\Sigma_{\rm H_2}}
=\frac{N_{\rm HI}}{2N_{\rm H_2}}
=\frac{\Xhi \Ihi}{2\Xco \Ico}.
\ee

The conversion factor for the HI line is given by  \cite{kraus1966}
\begin{equation}
    X_{\rm HI}=1.823 \times 10^{18}{\rm H cm^{-2}[K~ km~s^{-1}]^{-1}},
\end{equation}
assuming that the HI gas is optically thin, because the observed brightness temperature ($<$ several tens K) is sufficiently lower than the excitation temperature ($\sim 200$ K) measured in the GC \cite{pohl+2022}.
For the CO to \Htwo\ conversion, we adopt the empirical relation including the galacto-centric distance $R$, which considers the metallicity gradient (increasing toward GC) \cite{arimoto+1996},
\begin{equation}
    {\rm log}\frac{X_{\rm CO}}{X_{\rm CO,\odot}}=0.41\left(\frac{R-R_0}{r_e}\right).
\end{equation} 
Assuming a local conversion factor of 
$X_{\rm CO,\odot}=2.0\times 10^{20}$ \xcounit
\cite{bolatto+2013,sofue+2020xco} and scale radius of the Galactic disc $r_e=5.1$ kpc for $R_0=8.2$ kpc \cite{sofue2022},
we have
\begin{equation}
X_{\rm CO} ({\rm GC}) = 0.51\times 10^{20}\ {\rm H_2 \ cm^{-2} [K \ km\ s^{-1}]^{-1}}, 
\end{equation}  
so that
\be 
\frac{\Xhi}{2\Xco} = 1.79 \times 10^{-2},
\ee  
and 
\be
\eta_\Sigma=1.79 \times 10^{-2}\frac{\Ihi}{\Ico}.
\ee

\subsection{$\fsig$ distribution on the sky}

In Fig. \ref{mapwide} and \ref{maps} we show the distribution of $\fsig$ on the sky calculated using the \dev HI and \twco intensity maps.
The outline of highly molecular region with high $\fsig \ge 0.9$ nearly coincides with the high \twco intensity region.
However, the $\fsig$ value is almost saturated above $\sim 0.9$ regardless the clumpy distribution of the \twco intensity. 
  
Longitudinal variation of $\fsig$ along the Galactic plane is shown in the right panels of Fig. \ref{loncut}. 
The bottom-right panel from Nobeyama CO and ATCA HI data shows a plateaued (saturated) distribution of molecular fraction in the CMZ  as high as $\fsig \ge 0.9$ between longitudes $l=-1\deg.1$ and $+1\deg.8$ and by the clear-cut shoulders at these longitudes.
This high $\fsig$ region is isolated and distinguished from the surrounding milder $\fsig$ regions, as is clearly seen in the wide area plots in the top panel.  

Fig. \ref{latcut} shows a vertical variation of $\fsig$ at $l=+0\deg.6$ made from the Nobeyama CO and ATCA HI surveys. 
Molecular fraction is as high as $\fsig\sim 0.98$ near the Galactic plane, while it rapidly decreases with latitude, dropping to $\fsig\simeq 0.9$ at $b \sim \pm 0\deg.2 \ (\pm 28 \epc)$, and $\sim 0.8$ at $b\sim \pm 0\deg.4 \ (56 \epc)$.
The peak position is slightly displaced from the galactic plane toward the south by $b\sim -0\deg.05$.

\begin{figure} 
\begin{center}    
\includegraphics[width=8cm]{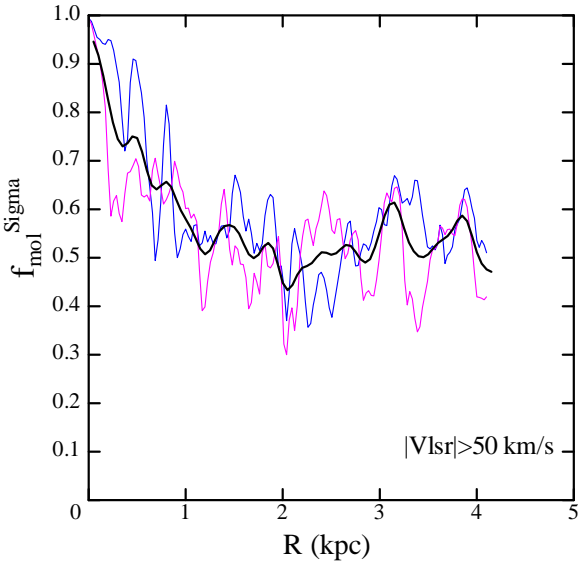} 
\end{center}
\caption{$\fsig$ from the Columbia CO and HI4PI data as a function of the galacto-centric distance $R$ as calculated from the longitude plot in Fig. \ref{mapwide}. Blue, pink and black lines represent curves at $l\ge 0\deg$, $<0 \deg$, and their average respectively.}
\label{FsigRkpc}
\end{figure}

Fig. \ref{FsigRkpc} shows $\fsig$ plotted against Galacto-centric distance $R=8 \sin\ l$ (kpc) as calculated using top-right panel of Fig. \ref{loncut}.
The plot is consistent with the radial plots of $\fmol$ obtained from the same data but using different methods \cite{koda+2016,sofue+2016}.
However, the present result shows lower values $\sim 0.5-0.6$ beyond $R\sim 2$ kpc.
This is because the present method measures the mean along the line of sight on the sky including larger radii than the plotted radius.
It is also interesting to note the local maxim corresponding to the the 3-kpc expanding arm/ring and the 4-kpc molecular ring 

\subsection{Volume molecular fraction $\frho$ from LVD} 

In order to examine the molecular fraction in the molecular Arms and clouds, we make use of longitude-velocity diagrams (LVDs) of HI and CO brightness temperatures $\Tb$, from which we calculate the volume-density molecular fraction (hereafter, volume molecular fraction).
LVDs were constructed using the \twco and HI cube data smoothed between $b=-0\deg.1$ and $+0\deg.1$ in order to cover the GC Arms I and II as well as the molecular complex around Sgr B2.
Fig. \ref{lvd} shows the obtained LVDs in HI, CO, their overlay, and volume molecular fraction.

The molecular Arms and clouds are now clearly visible and resolved, which were not recognized in the integrated intensity maps on the sky. 
Prominent features are the bright CO emitting region around Sgr B2 and B1 at $(l,\vlsr)\sim (0\deg.5, 60\ekms)$, the tilted LV ridge representing the molecular Arms running from $(l,\vlsr)\sim (1\deg.5,120 \ekms) $ to $(-1\deg,-200 \ekms)$, and the oval or parallelogram showing the expanding molecular ring or cylinder (EMR or EMC) with high non-circular motion of $\pm 150 \ekms$.

The volume molecular fraction is written in terms of the brightness temperature, $\Tb$, instead of the integrated intensity $I$ as follows:
\begin{equation}
    \frho
    \red{=\frac{\rho_{\rm H_2}}{\rho_{\rm HI} + \rho_{\rm H_2}} }
=\frac{1}{1+\eta_\rho}.
\end{equation}
and
\be
\eta_\rho=\frac{\rho_{\rm HI}}{\rho_{\rm H_2}}
=\frac{n_{\rm HI}}{2n_{\rm H_2}},
\ee
where $\rho$ and $n$ denote the volume mass density and number density of the gas, respectively.
Let us recall that the gas density is related to the column density and brightness temperature through
\begin{equation}
    n_i
    =\frac{dN_i}{dx}
    =\frac{dN_i}{dv}\frac{dv}{dx}    
    \red{=\frac{d (X_i\int T_{b,i} dv)}{dv}\frac{dv}{dx}}
        =X_i {\Tb}_i \frac{dv}{dx},
\end{equation} 
where $X_i$ is constant with $i=$ CO or HI.
Then, the HI to \Htwo\ mass ratio is approximated by
\begin{equation}
    \eta_\rho=\frac{n_{\rm HI}}{2n_{\rm H_2}} 
    \simeq  1.79\times 10^{-2} \frac{{\Tb}_{\rm HI}}{{\Tb}_{\rm CO}} \frac{[dv/dx]_{\rm HI}}{[dv/dx]_{\rm CO}}.
\end{equation}  
The radial velocity of HI and CO can be written as
\begin{equation}
    v_i(x)={\vrot}_i \sin \theta _i +\delta v_i,
\end{equation}
where $i=$ HI or CO, ${\vrot}_i$ is the rotation velocity, $\delta v_i$ is the velocity dispersion, and $\theta_i$ is the angle of the gaseous motion and line of sight.
We may safely assume that the rotation velocities of HI and CO are equal and the velocity dispersion is much smaller than $\vrot$ and almost constant along the line of sight.
Thus, we have 
\be
\frac{dv}{dx}({\rm HI})\simeq \frac{dv}{dx}({\rm CO}),
\ee or
\begin{equation}
    \eta_\rho\simeq  1.79\times 10^{-2} \frac{{\Tb}_{\rm HI}}{{\Tb}_{\rm CO}}.
\end{equation}  
This formula is used to calculate the distribution of $\frho$ in the LVD.

\subsection{$\frho$ distribution in the LVD}

\begin{figure*} 
\begin{center}   
\includegraphics[width=15cm]{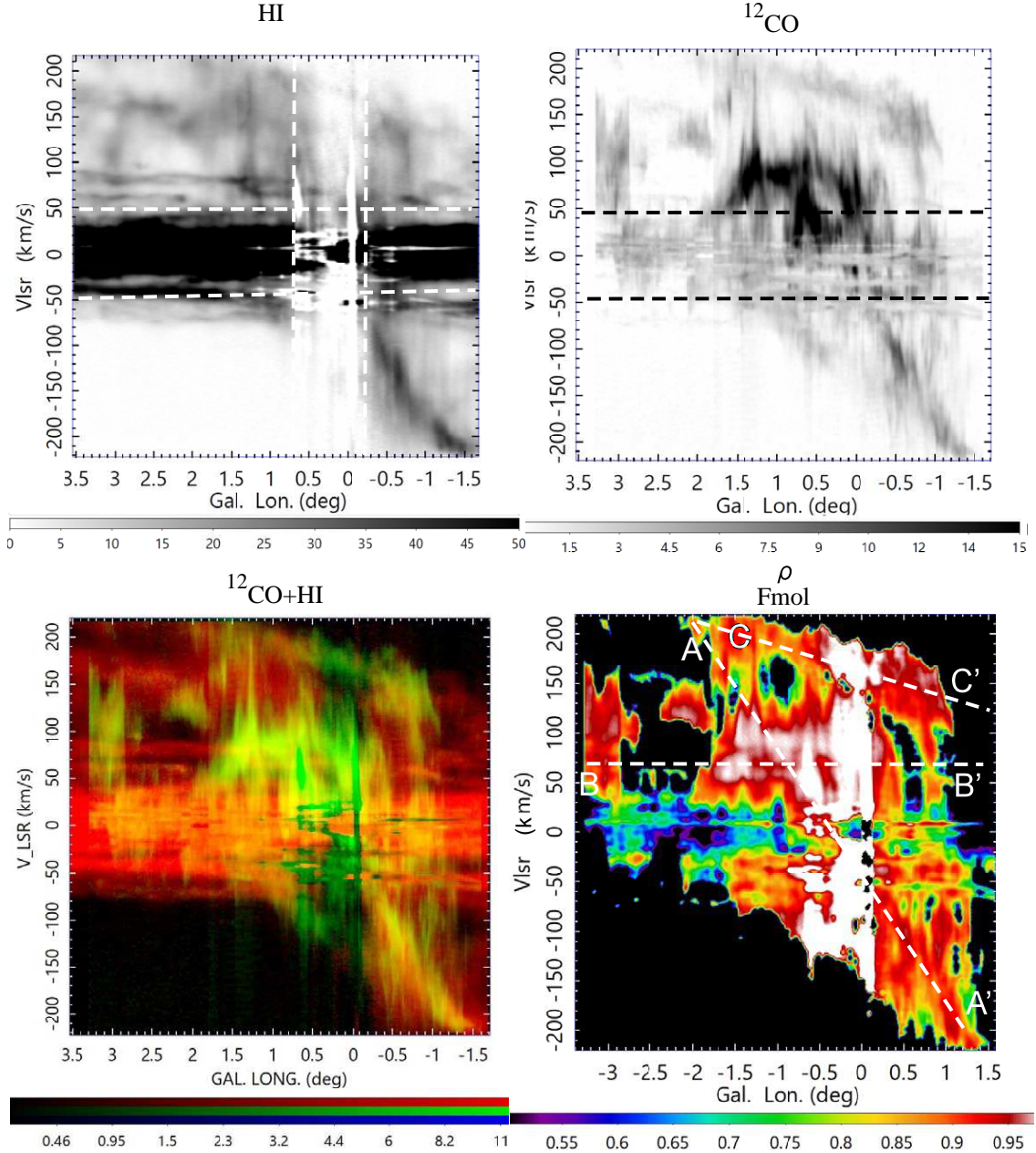}  
\end{center}
\caption{[Top left] LVDs of HI $\Tb$ (in K). Horizontal lines enclose the velocity range for DEV50, and vertical lines show longitude range with significant absorption against continuum.
[Top right] \twco $\Tb$ (in K) smoothed between $b=-0\deg.1$ and $+0\deg.1$ (width $0\deg.2=12'=29$ pc). 
[Bottom left] Overlay of \twco\ in green (0 to 15 K in Asinh scaling) on HI in red (0 to 50 K Asinh). 
[Bottom right] Volume molecular fraction, $\frho$, smoothed to the same resolution as that of HI map ($145''$). 
} 
\label{lvd}
\end{figure*}

As shown in the $\Tb$ LV diagrams in Fig. \ref{lvd}, the HI and CO temperatures are comparable to each other except for the region with significant HI absorption at low velocities and near Sgr A and B.
\red{
This means that the CMZ is composed mainly of molecular gas, 
because equal $\Tb$ between HI and CO means $\eta \simeq 1.79\times 10^{-2}$, or $\frho\simeq 1/(1+0.0155)=0.989$, so that  the molecular fraction is accordingly close to unity.}

In the bottom panel of Fig. \ref{lvd} we show the $\frho$ distribution as calculated using the top and middle panels for $\Tb$. 
The plot shows that $\fmol$ increases toward the density peak of the molecular arm, where $\fmol$ attains a maximum as high as $\simeq 0.95$, while it decreases in the enveloping region to $\fmol \sim 0.8$. 
The fraction attains the maximum toward the molecular complex of Sgr B2, where it is as high as $\frho\simeq 0.98-0.99$. 
Here, the data have been smoothed in the longitude direction to the HI resolution of 145$''$.
Note that CO $\Tb$ in the Sgr B2 region is higher than HI by twice, apparently indicating $\eta\sim 0.005$ and $\fmol\sim 0.995$.
But such extreme values are due to suppressed HI temperature by absorption against continuum emission of Sgr B as well as to self absorption, and hence do not represent true $\frho$.  
 
\begin{figure} 
\begin{center}   
 \includegraphics[width=7cm]{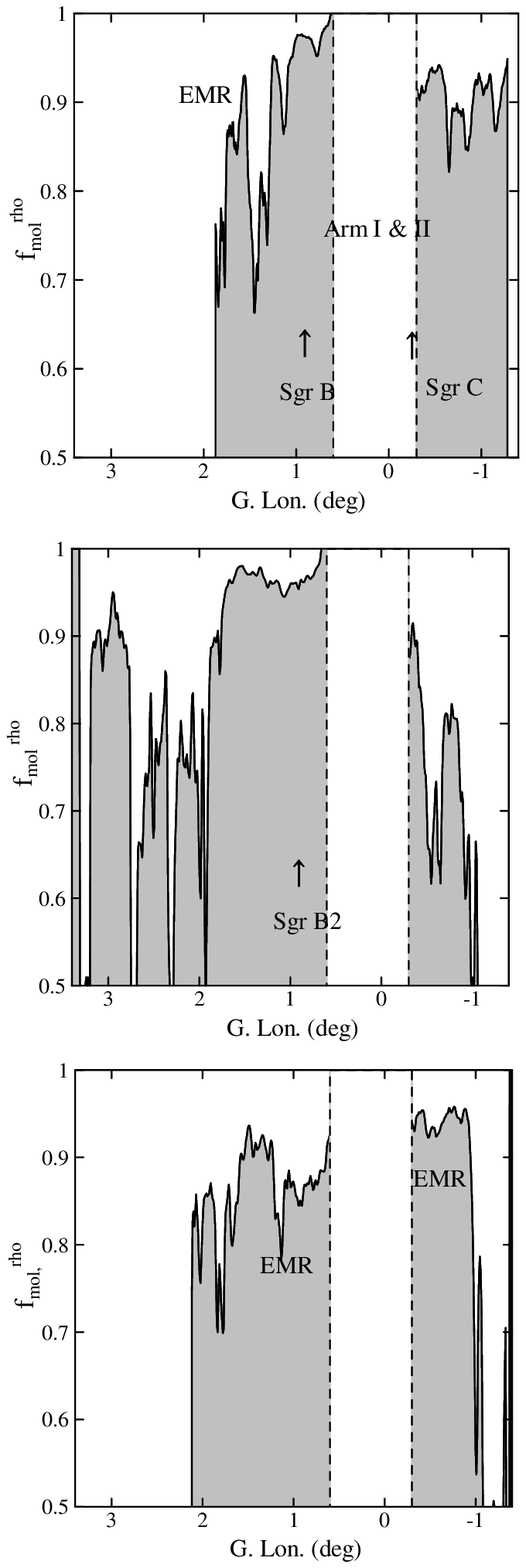} 
\end{center}
\caption{[Top] $\frho$ variation from A to A' of Fig. \ref{lvd} along GC Arms I and II from $(l,\vlsr)=(+1\deg.8, +220 \ekms)$ to $(-1\deg.2, -220 \ekms)$. 
[Middle] B to B' across Sgr B2 at $\vlsr=+62 \ekms$.
 [Bottom] C to C' along the expanding molecular ring (EMR) (or parallelogram) from $(l,\vlsr)=(+2\deg, +220 \ekms)$ to $(-1\deg, +150 \ekms)$. 
} 
\label{lvcut}
\end{figure}

Fig. \ref{lvcut} shows cross sections of LVD along the three lines A-A', B-B', and C-C' in Fig. \ref{lvd}, which represent $\frho$ variation along the GC Arms, across the molecular complex of Sgr B, and across the expanding molecular ring, respectively.
We stress that the molecular fraction along A-A' and B-B' are almost saturated at high level of $\frho\ge 0.96-0.98$.
{\red On the other hand, EMR/EMC exhibits a slightly milder $\frho$ of $\sim 0.9-0.95$ than the Arms, while it has an order of magnitude less density \cite{sofue2022}.}

\subsection{Dependence on the conversion factor}

\begin{figure} 
\begin{center}  
 \includegraphics[width=8cm]{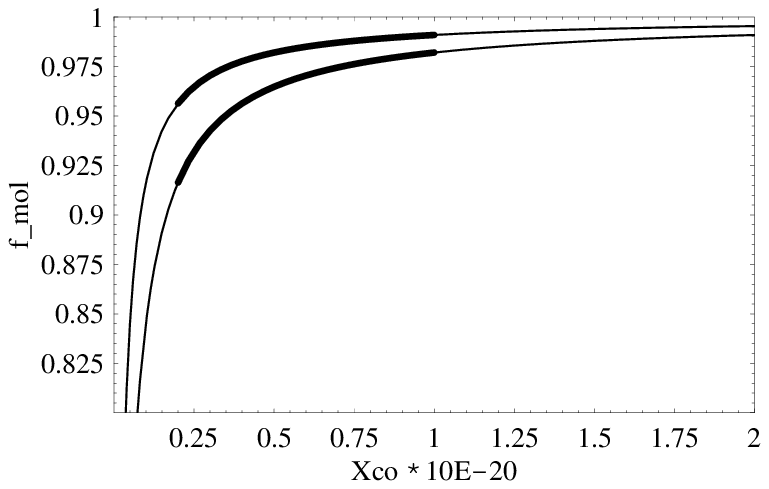} 
\end{center}
\caption{\red{Dependence of the molecular fraction on the $\Xco$ factor for the case with $\Ihi/\Ico=1$ (lower line) and 0.5 (more CO, upper line) (or $\Thi/\Tco=1$ and 0.5), where $\Xco$ in \xcounit.}
} 
\label{fmol-xco}
\end{figure}

\red{The \twco-to-\Htwo conversion factor adopted here, $\Xco=0.51\times 10^{20}$ \xcounit, is about a quarter of the local value of the Galactic molecular clouds.
This is consistent with the molecular mass estimations in the GC obtained by various methods other than CO \cite{dahmen+1998}.
An even smaller value of 0.24 \xcounit for the \twco($2-1$) line, which has about the same intensity as \twco($1-0$), has been also obtained \cite{oka+1998b}.
Therefore, $\Xco$ may have an uncertainty of a factor of two that affects the molecular fraction.
Figure \ref{fmol-xco} shows the dependence of $\fmol$ on $\Xco$ for $\Ihi/\Ico=1$ and 0.5 or $\Thi/\Tco=1$ and 0.5, which represents typical cases of CMZ.
A range of $\fmol$ between $\Xco \sim 0.25$ and $\sim 1$, as indicated by the thick line, will be still acceptable due to our current uncertain knowledge of the conversion factor in the GC.
However, even taking into account such uncertainties, the definition of the CMZ based on the relative variation of $\fmol$ in the GC region will not change.}

\section{Discussion: How central and molecular is CMZ?} 

\subsection{Comparison with interstellar clouds and disc gas}

The presently detected high $\fmol$ and low $\eta$ in the CMZ may be compared with the values for interstellar clouds such as observed in $\rho$ Oph with 
$\eta_\Sigma=N_{\rm HI}/(2N_{\rm H_2})=1/112 \ (\fsig = 0.99)$
\cite{minn1981}, 
 $\eta_\Sigma\sim 10^{-2}\ (\fsig\sim 0.99)$ of a GMC \cite{sato+1992},
or $\eta_\Sigma\sim 0.1-0.15 \ (\fsig\sim 0.85-0.9)$ of Sgr B2 \cite{lang+2010}.
However, the highest value observed toward Sgr B, $\frho=0.99$, in this study is still lower than that obtained in a local dark cloud having an extremely high value of 
$\eta_\Sigma \sim 5 -27\times 10^{-4} \ (\fsig\sim  0.997-0.999)$ \cite{goldsmith+2005}.
On the other hand, the CMZ's $\fsig$ is far greater than the values observed in HI clouds as $\fsig\sim 0.2-0.5$ \cite{nak+2020}, and those along sight lines toward prominent radio sources through the arms ($\fsig \sim 0.7-0.9$) and inter-arm regions ($\sim 0.1-0.3$) \cite{winkel+2017}.

\subsection{Definition of CMZ: Isolated island}

Using the spatial distribution of the surface-mass molecular fraction in Fig. \ref{maps} and \ref{loncut}, we define the CMZ as a region with a plateaued distribution of high molecular fraction \red{between the steeply increasing gradients in both sides of the center at $\fsig \sim 0.8-0.9$.}
The region is well defined between the edges at $l=-1\deg.1 \ (-157 \epc)$ and $+1\deg.8\ \ (+257\epc)$ with eye-estimate errors $\pm 0\deg.05$, or the scale radii to be $r_{-}=-157$ and $r_{+}=257$ pc ($\pm 7$ pc).
It is stressed that the positions of the density shoulders exactly coincide with those of $\fsig$.

Namely, the CMZ is a region with plateaued distribution of \Htwo\ gas having shoulders both in the column-density (intensity) and molecular fraction profiles.
So, this is the answer to the question of how much central and molecular the CMZ is.
The measured extent of CMZ is thus consistent with that originally introduced \cite{morris+1996}.
Also, the plateaued nature indicates that the gas distribution inside CMZ is ring like \cite{henshaw+2022}.
A new point to be emphasized by this analysis is that the CMZ is an isolated island in the shallow sea of low density HI that has not sufficient mass to supply the CMZ.

\begin{figure} 
\begin{center}    
\includegraphics[width=7cm]{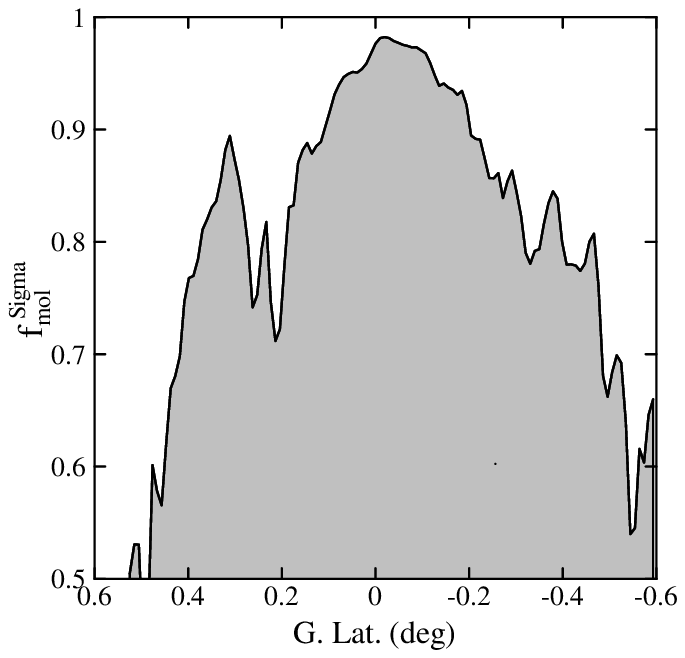} 
\end{center}
\caption{Latitudinal (vertical) variation of $\fsig$ across the Galactic plane at $l=0\deg.6$.}
\label{latcut}
\end{figure}

The latitudinal profile of $\fsig$ is shown in Fig. \ref{latcut}, which is well represented by a Gaussian profile.
The intensity distribution also has a Gaussian profile in the latitudinal direction \cite{sofue2022}.
From Fig. \ref{latcut}, the latitudinal extent at a threshold level of $\fsig=0.9$ is measured to be $z_{\rm 0.9}=0\deg.2$ with eye estimate error of $ \pm \sim 0\deg.02$.
This height is about equal to the column-density's scale half thickness
.
So, we define the latitudinal extent of the CMZ by this height, $h_z=\pm 0\deg.2$ ($\pm 29$ pc).

Thus defined CMZ is significantly asymmetric, or lopsided, with respect to Sgr A$^*$ (nucleus), being displaced to positive longitude by $\delta l \sim 0\deg.35 \ (50 \epc)$.
It is also offset from the Galactic plane toward the south by $\delta b\sim -0\deg.05 \ (7 \epc)$.

It is impressive that the CMZ is highly isolated from the surrounding less molecular region, particularly in the wider area plot in Fig. \ref{loncut} (b) and (d), making an island of molecular gas in the central $\sim 430 \epc$. 
The CMZ composes a distinguished galactic structure (object) with the highest molecular fraction in the Milky Way.
Despite of the high $\fmol$ inside, it is surrounded by a low $\fmol$ region or a gap with $\fsig\sim 0.5-0.7$ at $l>\sim 2\deg$ and $<\sim -1\deg.2$ (Fig. \ref{loncut}).  
However, the current studies of the Milky Way using the Columbia CO survey \cite{dame+2001} and HI survey using the Parkes 64-m telescope survey \cite{HI4PI+2016} at lower resolutions have shown a monotonic increase of $\fmol$ toward the GC inside $R\le \sim 2$ kpc. 
Studies of external spiral galaxies have also shown a monotonic increase of $\fmol$ toward the centers 
\cite{sofue+1995fmol,honma+1995,kuno+1995,hidaka+2002,nak+2006,tosaki+2011,tanaka+2014,sofue+2016,koda+2016}. 

The variable $\fmol$ outside CMZ has been detected here thanks to the high resolution ($\sim 6 $ pc), while the current works had effective resolutions of $\sim 100$ pc or larger, so that the innermost large variation of $\fmol$ has not been detected.
It is an interesting question whether the $\fmol$-gap around the CMZ is a general phenomenon, or it is specific in the Milky Way's center.
This will be a subject for higher resolution observations in the CO and HI lines of spiral galaxies in the future.

\subsection{Volume-filling factor and the difference of $\frho$ and $\fsig$}

The volume-mass molecular fraction was observed to be generally higher than the column-mass molecular fraction, or $\frho > \fsig$, in the CMZ.
From Fig. \ref{loncut} and \ref{lvcut} we approximate the mean values to be  $\frho\sim 0.96$ and  $\fsig \sim 0.93$.
The lower $\fsig$ than $\frho$ is attributed to the volume filling factor $\alpha_{\rm H_2}$ of \Htwo\ and $\alpha_{\rm HI}$ of HI gas.

We here assume that each of molecular and HI clouds has a nearly constant density and write the column densities as $N_{\rm HI}\sim \langle \rho_{\rm HI} \langle $ and
$2N_{\rm H_2}=\langle \rho_{\rm H_z} \rangle$, where the angle-parentheses denote the average along the line of sight.
We here introduce filling factors $\alpha_{\rm HI}$ and $\alpha_{\rm H_2}$ for HI and molecular clouds and rewrite the average in terms of the filling factors as
$\langle \rho_{\rm HI}\rangle=\alpha_{\rm HI} \rho_{\rm HI}$ 
and 
$\langle \rho_{\rm H_2}\rangle=\alpha_{\rm H_2}\rho_{\rm H_2}$.
Recalling that $\eta_i \ll 1$, we obtain
\be
\fsig=\frac{1}{1+\eta_\Sigma} 
\simeq 1-\frac{\langle \rho_{\rm HI}\rangle}{\langle \rho_{\rm H_2}\rangle}
\simeq 1-\frac{\alpha_{\rm HI}\rho_{\rm HI}}{\alpha_{\rm H_2} \rho_{\rm H_2}}.
\ee
On the other hand we have
\be
\frho
\simeq 1-\frac{\rho_{\rm HI}}{\rho_{\rm H_2} }.
\ee
So, we obtain
\be
\Delta f=\frho-\fsig 
\sim \left(\frac{\alpha_{\rm H_2}}{\alpha_{\rm HI}} -1 \right)\frac{\rho_{\rm HI}}{\rho_{\rm H_2}}.
\ee

The observed relation 
$\frho>\fsig$ indicates 
$\alpha_{\rm H_2}>\alpha_{\rm HI}$.
If we assume 
$\rho_{\rm HI}/\rho_{\rm H_2}\sim 0.01$
(e.g., 
$n_{\rm HI}\sim 20 \ {\rm H \ cm^{-3}},$
$n_{\rm H_2} \sim 10^3 \ {\rm \ H_2 \ cm^{-3}}$
) for example, the observed 
$\Delta f\sim 0.03$ 
yields 
$\alpha_{\rm H_2}/\alpha_{\rm HI}\sim 2$.
The CMZ is thus a molecular gas dominant region not only by mass and density, but also by the volume filling factor.

\subsection{Origin of high molecular fraction in CMZ}

The molecular fraction is determined by the gas pressure $P$, metallicity $Z$, and ultra-violet (UV) radiation field $U$ of the interstellar gas \cite{elme1993}.
It can be expressed by a function of the galacto-centric distance $R$, given a Galactic disc model for the radial variation of $P, Z$ and $U$.
While $U$ suppresses $\fmol$, $P$ (density) and $Z$ act to increase $\fmol$, where $Z$ doubly affects to increase $\fmol$ by UV shielding and catalystic formation of \Htwo\ on dust grain surfaces.
\red{It has been shown that the thus calculated $\fmol$ monotonically increases toward the GC with a scale radius comparable to that of the three quantities of about $R_e\sim 5-7$ kpc \cite{honma+1995,tanaka+2014,sofue+2016,koda+2016}.}

In so far as the exponential variations are assumed for $P$, $U$ and $Z$, as in the current works, any model results in a monotonic increase of $\fmol$ toward the GC.
So, the observed rapid $\fmol$ variation at $R\sim 200$ pc must be due to more local variation of the three quantities.
Among the three parameters, the metallicity $Z$ cannot be variable at scale radius less than $\sim 1$ kpc, because the time scale of variation is on the order of the Galaxy's evolution time including the Galactic bulge. 
The radiation field $U$ is expected to be highest inside CMZ according to the wealth of thermal radio continuum sources indicating a number of HII regions \cite{sofue1990,sofue2022,yz+2022}.
This means that the UV field cannot be the cause for depression of $\fmol$ beyond $R\sim 200$ pc.

Therefore, only the parameter that can affect the sudden change in $\fmol$ at $R\sim 100-200$ pc is the pressure $P$ which is proportional to $\rho T$ with $\rho$ and $T$ being the ISM density and temperature. 
Since the UV field is weaker outside CMZ, the temperature may be lower outside CMZ, which means that the pressure equilibrium between outside and inside CMZ does not hold.
So, only one possible mechanism for $P$ to be increased locally inside CMZ is the dynamical confinement of ISM in the deep Galactic potential.

Such dynamical confinement can be obtained by accretion of the gas in the surrounding disc, where $\fsig  \sim 0.5$ so that the gas is half molecular and half in the shallow HI envelope (CHZ), onto the central region by the loss of angular momentum by galactic shock waves in the bar potential \cite{sorensen+1976,roberts+1979,wada+1995}. 
The gas between the center and bar ends are swept away to make low gas density region surrounding the accretion disc.   

Once it is accumulated in the center, the gas is converted to molecules by the phase transition due to the increase in the density ($\propto P$) and maybe also due to locally high metallicity ($Z$) in the core of the Galactic bulge, unless the radiation field ($U$) is too strong such as in a starburst phase.
The accumulating dense gas composes there a ring in circular rotation as shown by the extensive numerical simulations \cite{li+2015,sormani+2019,tress+2020}.
The ring will be sustained for a long time without collapsing by their self-gravity because of the high differential rotation on the order of $\sim 1000\ {\rm kms\ s^{-1}\ kpc^{-1}}$ in the deep gravitational potential inside CMZ \cite{sofue2013}.

However, there remains a question whether the observed high $\fmol$ can be maintained still in such a high shear motion even though it is anticipated by the current $f_{\rm mol}$ theories that consider $P,\ Z$ and $U$, but do not take into account the dynamics.
In fact, the numerical simulations as above assuming a bar potential in the GC have shown rapid inflow of gas yielding a turbulent gas ring, and accordingly indicate a very low molecular fraction of $\fmol \sim 0.55$ \cite{tress+2020}. 

\subsection{No HI-to-\Htwo\ transition in the CMZ}

In such a high $\frho$ region as CMZ, the molecular Arms I and II, including Sgr B complex, cannot be produced by the phase change of surrounding HI gas by compression, because the HI reservoir inside CMZ contains far less mass than the arms.
Namely, the HI-to-\Htwo phase transition is not the major process to form the molecular gas structures in the GC.
So, the arms can be formed by coagulation of smaller molecular clouds and diffuse \Htwo\ by compression inside the CMZ \cite{koda+2016}.  

\subsection{Central HI Zone (CHZ)} 

We finally mention about the HI gas distribution from the DEV50 HI maps. The horizontal HI intensity distribution in Fig. \ref{loncut} shows enrichment of HI gas in GC. In contrast to the CMZ with sharp edges, the HI concentration is milder, continuous from the periphery, and shows a nearly linear increase towards the center in the semi-log plot.

Considering the absorption near Sgr A, the profile may be approximated by an exponential function with gradient of $e$ folding radius of $r_{\rm HI} \sim +3 \deg \ (\sim 430 \epc) $ on the positive longitude side, and $ \sim -2 \deg (\sim 285 \epc) $ on the negative longitude side. This "central HI zone (CHZ)" may constitute an extended outskirt of the CMZ. The CMZ stands as an isolated island in the extended shallow sea of HI.However, the CHZ does not work as a reservoir to supply the gas to CMZ because the density and total mass is too small \cite{sofue2022}.

More global HI concentration in the GC is visible in the DEV50 HI map as the peak in the central few degrees over the global minimum (Fig. \ref{loncut}(a), Fig. \ref{HIsimob}). Such HI enhancement in integrated intensity maps has been not clearly seen in the current presentations, being buried by the fore- and background disc. In the appendix, we demonstrate how efficient the DEV method is to eliminate the HI disc contamination and to quantify the density distribution of HI gas in the CHZ.

We point out that the longitudinal HI distribution is asymmetric with respect to the center in the sense that CHZ is shifted to the west (negative longitude side) by $\sim 1-2\deg$. Interestingly, the HI asymmetry is in the opposite sense to that of the CMZ that is shifted to the east (positive longitudes).

\section{Summary}

We have obtained a detailed comparison of the CO and HI maps and LV diagrams in the GC region, and showed that the molecular fraction is as high as $\frho \sim 0.9-0.95$ in the CMZ.
The highest fraction was observed toward Sgr B2 molecular complex at $\frho\sim 0.98$, and it remains at high values $0.93-0.98$ in the GC Arms I and II.
\red{The expanding molecular ring or the parallelogram shows a slightly milder fraction of $\frho\sim 0.9-0.93$, but with an order of magnitude lower gas density than CMZ.
We quantitatively re-defined the CMZ's location and extent as the region on the sky with the column (projected) molecular fraction at $\fsig \ge 0.8-0.9$ spanning from $l=-1\deg.1$ to $+1\deg.8$ and $b=-0\deg.2$ to $+0\deg.2$, and answered the question how much central and molecular the CMZ is. 
We also defined the Central HI zone (CHZ) as an HI enhanced disc from $-2\deg$ to $+2\deg.5$ and $|b|\le 0\deg.5$, which envelopes the CMZ.}

\vskip 3mm
\noindent{\bf ACKNOWLEDGEMENTS}
\vskip 1mm

\noindent The author express his sincere thanks to the authors and collaborators of the archival data as follows. Data analysis was performed at the Astronomy Data Center of the National Astronomical Observatory of Japan.
\vskip 2mm

\noindent{\bf DATA AVAILABILITY}
\vskip 1mm

\noindent The CO- and HI-line data were downloaded from  \\
https://www.nro.nao.ac.jp/~nro45mrt/html/results/data.html \\ 
https://www.atnf.csiro.au/research/HI/sgps/GalacticCenter/ 
Home.html.
\vskip 2mm

\noindent{\bf NO CONFLICTS OF INTEREST} 
\vskip 1mm

\noindent I have no conflicts of interest.


\begin{appendix}

\section{CO and HI maps}

We reproduce the \twco and HI intensity maps used to derive the DEV50 maps from \cite{sofue2022}.

\begin{figure*} 
\begin{center}    
\includegraphics[width=150mm]{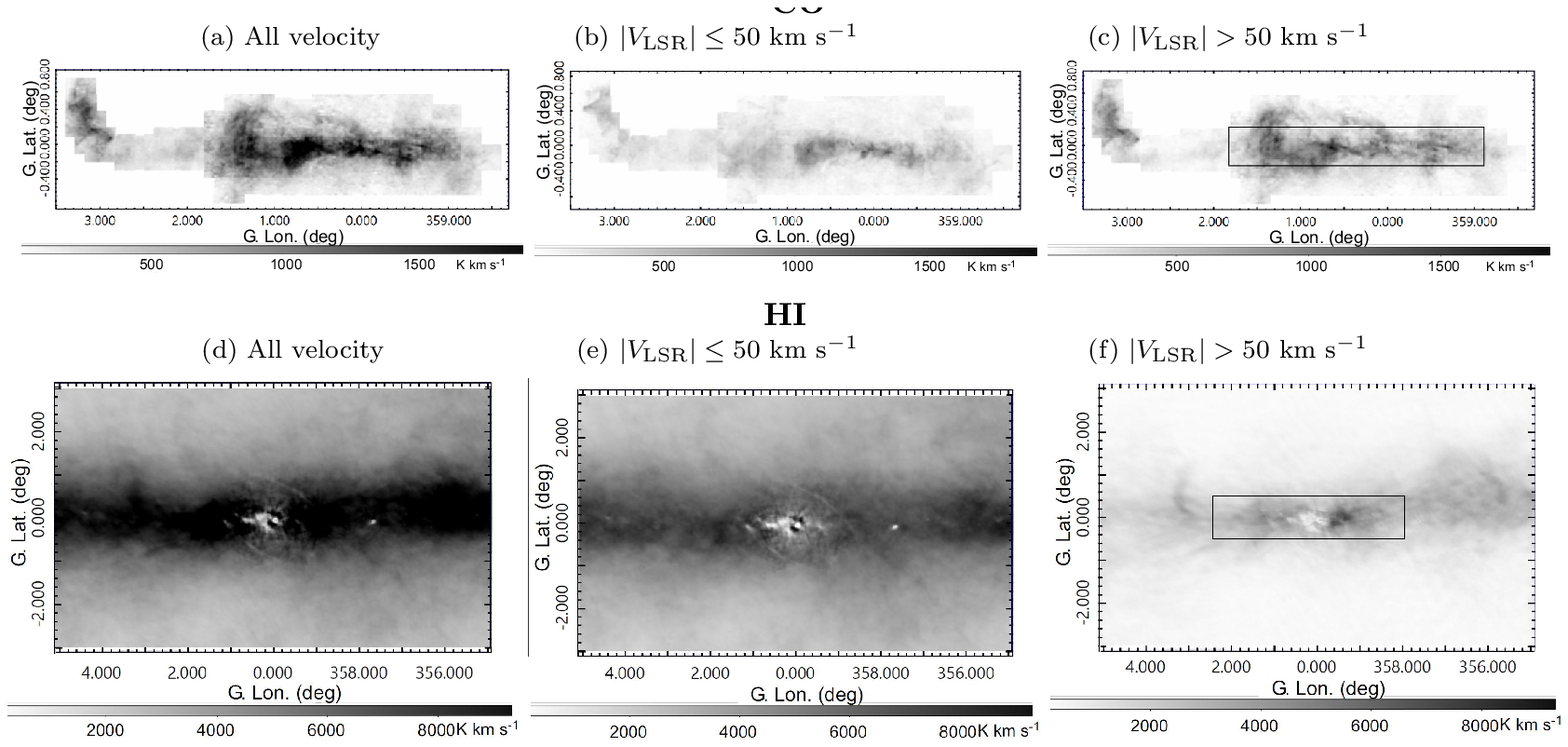}  
\end{center}
\caption{
Integrated intensity maps using the Nobeyama $^{12}$CO (Oka et al. 1998a) and ATC HI (McClure-Griffiths et al. 2012) surveys, as reproduced from our previous paper (Sofue 2022). Squares in the top and bottom panels mark the CMZ and CHZ, respectively
}
\label{mapA}
\end{figure*}

\section{DEV50 HI longitudinal profile}

\begin{figure} 
\begin{center}    
\includegraphics[width=7cm]{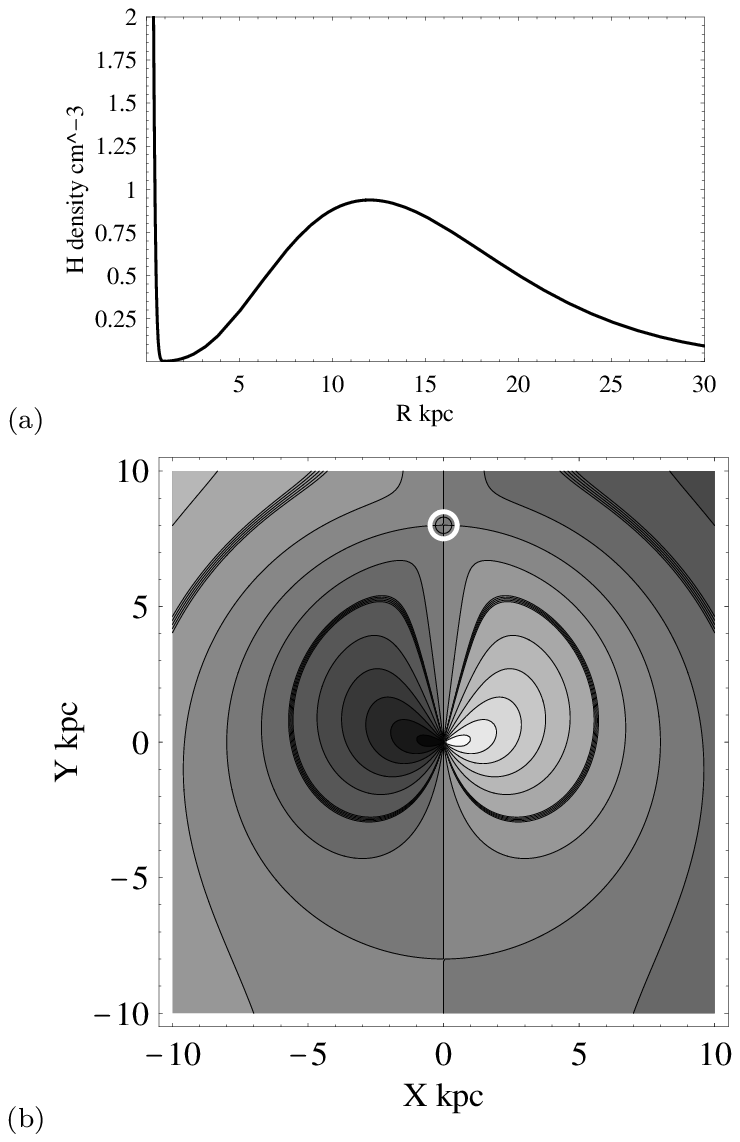}  
\end{center}
\caption{(a) Model radial distribution of HI density in the Galactic plane.
(b) Velocity field for flat rotation curve with $\vrot=200 \ekms$. The Sun is at the white circle. DEV50 integration is obtained outside the fan-shaped area with $|\vlsr|> 50 \ekms$ as indicated by the thick contours.
Contours are drawn at interval of 25 \kms, with the Solar circle at $\vlsr=0$ \kms and the thick contours are $\pm 50$ \kms.}
\label{model}
\end{figure}

We check how well the DEV50 profile for a HI gas distribution model can reproduce the observed profile. 
We assume a model density distribution as follows, composed of a disc mimicking the observed HI disc \cite{nak+2016} and an HI central disc representing the CHZ, respectively,
\be \rho=\rho_1+\rho_2,\ee
where 
\be \rho_1=0.2 (r/r_{\rm d})^4 \exp(-(r/r_{\rm d}))\ee
represents the disc component and
\be \rho_{\rm c}=10 \exp(-(r/r_{\rm c})^2)\ee
is the CHZ. The scale radii are taken to be
$r_{\rm d}=3$ kpc and $r_{\rm c}=0.3$ kpc.
We also assume a simple rotation cuve with constant velocity $\vrot=200 \ekms$.
Fig. \ref{model} shows the model density distribution and radial-velocity field.

\begin{figure} 
\begin{center}      
\includegraphics[width=8cm]{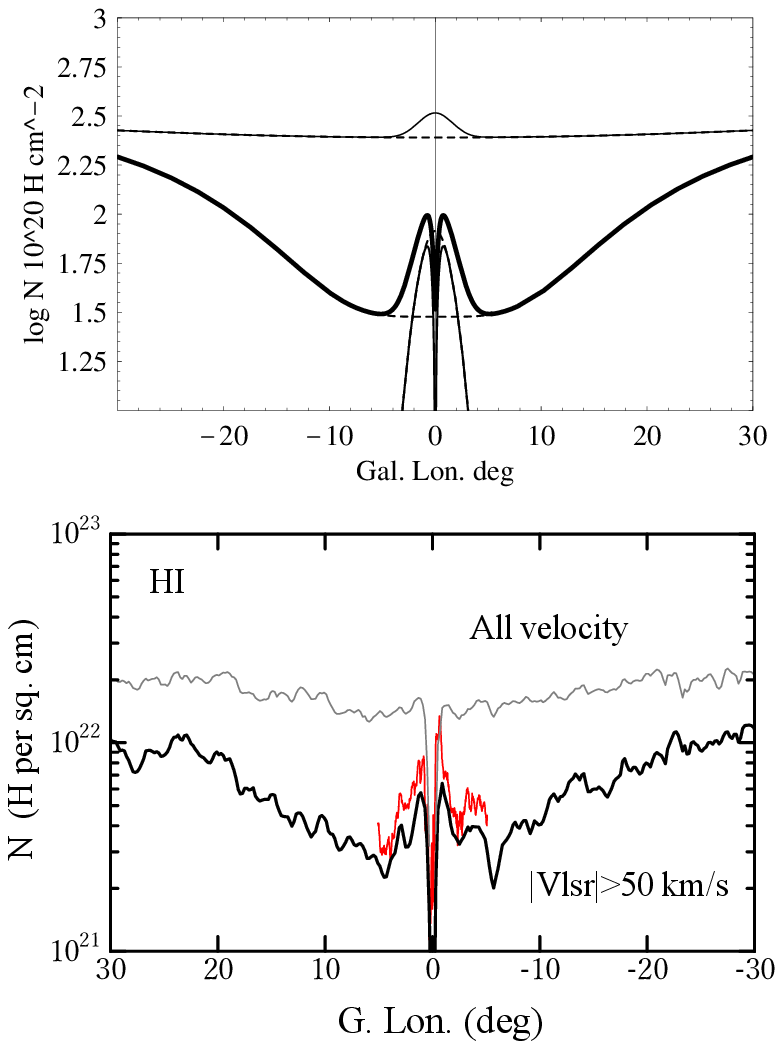} 
\end{center}
\caption{ (a) Longitudinal profile of column density $N$ of HI by DEV50 (thick line) for a flat rotation curve with $\vrot=200 \ekms$, and non-DEV (thin line).
(b) Observed HI column density by DEV50 (thick line) and non-DEV using HI4PI HI data. Red line shows DEV50 profile for ATCA HI data.}
\label{HIsimob}
\end{figure}

Calculated longitudinal profile of the HI column density is shown in Fig. \ref{HIsimob}.  
The thick line shows the result for DEV50, and the upper thin line shows the profile without DEV.
The bottom panel shows the observed column density profiles for DEV50 by thick line and non-DEV integration (thin line) for the HI4PI survey along the Galactic plane.
The red line is the DEV50 profile using the ATCA survey, which shows about 50\% larger densities.
\red{The higher density in ATCA profile may be due to the higher resolution more effectively picking up high-density structures near the Galactic plane as well as to smaller absorption effect against the continuum sources compared to single-dish observations by HI4PI.}

\begin{figure} 
\begin{center}     
\includegraphics[width=8cm]{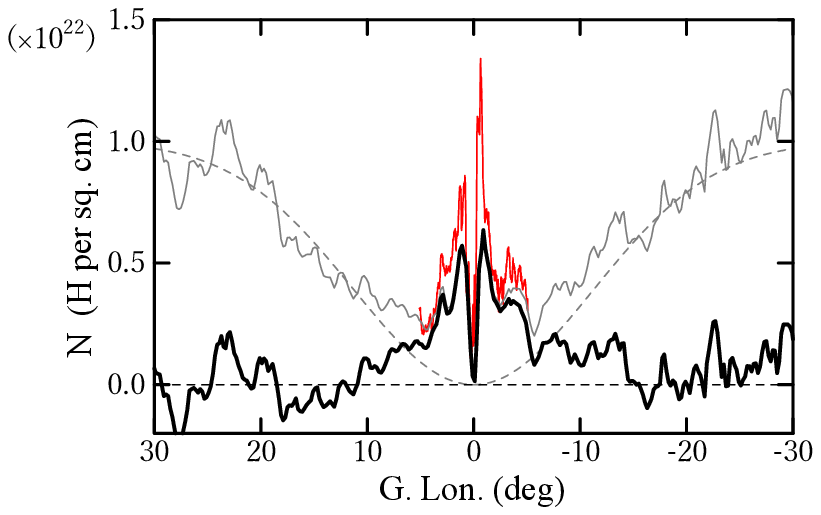} 
\end{center}
\caption{$N$ of HI gas (thick line) in linear scaling after subtracting the residual disc component, as fitted by a Gaussian profile (dashed line) to the observed DEV50 $N$ map (grey line) from HI4PI HI data. Red line shows the same, but for ATCA HI data.}
\label{NHI-CHZ}
\end{figure}

The DEV50 longitudinal profile can be used to abstract the central HI zone. In Fig. \ref{NHI-CHZ} we show a profile (thick line) after subtracting the extended HI component by fitting to the outer profile (thin line) at $|l|\ge 15\deg$ by a Gaussian function with scale radius $16\deg$ and amplitude $1.0\times 10^{22}$ H cm$^{-2}$ (dashed line). The CHZ is clearly visible as the central concentration inside $-4\deg \le l \le +2\deg.5$, and is further surrounded by a broader enhancement from $l\sim -13\deg$ to $\sim +10\deg$. 

\end{appendix} 
\end{document}